\documentclass{aa}  

\usepackage{natbib}
\bibpunct{(}{)}{;}{a}{}{,}

\usepackage{graphicx}
\usepackage{txfonts}
\newcommand{\vai}{v_{\mathrm{A,i}}}

\newcommand{\rhoe}{\rho_{\mathrm{e}}}
\newcommand{\rhoi}{\rho_{\mathrm{i}}}

\begin{document}

\title{Prominence seismology using the period ratio\\ of transverse thread  oscillations}

\titlerunning{Prominence seismology using thread oscillations}

   \author{R. Soler\inst{1},  M. Goossens\inst{2}, \and J. L. Ballester\inst{1}}

   \institute{Departament de F\'isica, Universitat de les Illes Balears,
              E-07122 Palma de Mallorca, Spain\\
              \email{roberto.soler@uib.es}
         \and
            Centre for mathematical Plasma Astrophysics, Department of Mathematics, KU Leuven, Celestijnenlaan 200B, 3001 Leuven, Belgium \\
               \email{marcel.goossens@wis.kuleuven.be}
             }

   \date{Received XXX; accepted XXX}

 
  \abstract
   {The ratio of the period of the fundamental  mode to that of the first  overtone of kink oscillations, from here on the `period ratio', is a seismology tool that can be used to infer information about the spatial variation of density along solar magnetic flux tubes. The period ratio is 2 in longitudinally homogeneous thin tubes, but it differs from 2 due to longitudinal inhomogeneity. In this paper we investigate  the period ratio in longitudinally inhomogeneous prominence threads and explore its implications for prominence seismology. We  numerically solve the two-dimensional eigenvalue problem of kink oscillations in a model of a  prominence thread. We take into account three nonuniform density profiles along the thread. In agreement with previous works that used simple piecewise constant density profiles, we find that the period ratio is larger than 2 in prominence threads. When the ratio of the central density to that at the footpoints   is  fixed, the period ratio depends strongly on the form of the density profile along the thread. The more concentrated the dense prominence plasma near the center of the tube, the larger the period ratio.   However, the period ratio is found to be independent of the specific density profile  when the spatially averaged density in the thread is the same for all the profiles. An empirical fit of the dependence of the period ratio on the average density is given and  its use for prominence seismology is discussed. }


   \maketitle
%

\section{Introduction}

High resolution observations reveal that solar prominences are formed by a myriad of long and thin sub-structures usually called threads or fibrils. In observations of prominences above the limb, vertical threads are commonly seen in quiescent prominences \citep[e.g.,][]{berger2008}, while horizontal threads  are usually observed in active region prominences \citep[e.g.,][]{okamoto2007}. However, it has been suggested  \citep[e.g.,][]{schmieder2010} that vertical threads might actually be a pile up of horizontal threads which appear as vertical structures when projected on the plane of the sky. Along this line of thought, threads supported in quasi-horizontal magnetic fields seem to be more consistent  with  observations of filament threads on the solar disk \citep[see][]{lin2011}, with  determinations of the  magnetic field orientation in prominences \citep[e.g.,][]{casini2003,orozcofield2014}, and with  equilibrium models of prominences based on slightly dipped magnetic fields \citep[e.g.,][]{terradas2013}. In this work we use the term `thread' to refer to   prominence horizontal fine structures.

Transverse oscillations of prominence threads with periods roughly between 1 and 20 minutes have been frequently reported \citep[see, e.g.,][]{okamoto2007,lin2007,lin2009,ning2009,orozco2014}. From the theoretical point of view, the oscillations are usually interpreted as magnetohydrodynamic (MHD) kink modes of the flux tube that supports the prominence thread \citep[see, e.g.,][]{terradas2008,lin2009,soler2010,arregui2011,soler2011,soler2012}. Both observational and theoretical aspects of prominence thread oscillations have been reviewed by \citet{arregui2012}.

Prominence seismology  relies on the comparison of observed with predicted properties of prominence oscillations \citep[see][]{ballester2014}. The predicted properties are based on theoretical models. The aim is  to indirectly infer information about the plasma and/or the magnetic field in prominences \citep[see, e.g.,][]{terradas2008,lin2009,soler2010,arreguihinode2012}. In this direction, the ratio of the period of the fundamental longitudinal mode to period of the first longitudinal overtone of thread kink oscillations, from here on the `period ratio', can be used to obtain information about the spatial variation of density along the threads. Standing MHD waves on magnetic flux tubes can be characterised by the number of nodes in their eigenfunctions, and this classification is independent of the velocity polarisation of the wave. Modes can have different number of nodes in the longitudinal (axial) part of the eigenfunctions as well as in the radial part of the eigenfunctions. The terms fundamental longitudinal mode and first longitudinal overtone refer to waves that have no nodes and have only one node, respectively, in the longitudinal (axial) part of the eigenfunction. In the present work we study the fundamental longitudinal mode and the first longitudinal overtone of transverse kink waves. The period ratio is 2 in longitudinally homogeneous thin tubes, but it differs from 2 due to longitudinal inhomogeneity \citep{andries2005a,andries2005}. In the context of coronal loop transverse oscillations, the use of the period ratio as a seismology tool was first proposed by  \citet{andries2005} and has been exploited in a number of subsequent works \citep[e.g.,][among others]{mcewan2006,mcewan2008,dymova2006,donnelly2007,vandoorsselaere2007,verth2008,ruderman2008,arregui2013}. An extensive review of the use of the period ratio for coronal loop seismology can be found in \citet{andries2009}.  

Although no reliable simultaneous observations of the two periods in prominence threads are currently available, numerical simulations of thread oscillations by \citet{soler2011} indicate that the fundamental mode and the first overtone are the modes that are more easily excited in threads due to external disturbances. It is reasonable to expect that the two periods will be eventually detected in prominence threads by future high-resolution observations. Due to its great potentiality for seismology, a firm theoretical basis on the value of the period ratio in prominence threads is needed. The use of the period ratio for prominence seismology was first studied by \citet{diaz2010}. They considered a prominence thread model composed of a thin magnetic flux tube filled with a dense blob of  plasma that occupied a small part of the tube only, while the rest of the tube was evacuated, i.e., it was occupied by a much less dense plasma. The density in both the dense blob and in the evacuated part was homogeneous, and an abrupt jump between the two densities was assumed. Similar piecewise constant models have been used in previous works as, e.g., \citet{diaz2002,dymova2005,terradas2008,soler2010,soler2011,soler2012}, while a continuous but thin transition between the dense blob and the evacuated part of the tube was considered by \citet{arregui2011}. The simple piecewise constant  model used by \citet{diaz2010} allowed these authors to derive analytic expressions for the periods of the fundamental mode and the first overtone. They found that, contrary to the case of coronal loops where the period ratio is smaller than  2 due to stratification, in prominence threads the period ratio is larger than 2. The reason for this opposite result is that in prominence threads the densest plasma is located near the tube centre, while in stratified coronal loops the plasma is denser near the footpoints. Subsequently, \citet{soler2011} added longitudinal flow to the model of \citet{diaz2010} and found corrections to their results \citep[see also][]{robertus2014}.

 The purpose of this paper is to advance the study of the period ratio of kink oscillations in inhomogeneous prominence threads and to explore its  implications for prominence seismology. Here we go beyond the piecewise constant density profile used by \citet{diaz2010} and consider a more realistic continuous variation of density along the thread. Assuming that the thread formation is due to condensation of cool plasma via thermal instability \citep[see, e.g.,][]{luna2012}, the density profile along the threads is likely to depend on the energy balance between footpoint heating, radiative losses, and thermal conduction. This balance naturally leads to a spatially inhomogeneous density along the threads. Here, we evaluate the impact of the form of the longitudinal density profile on the period ratio. To do so, we numerically solve the full two-dimensional (2D) eigenvalue problem of kink oscillations in a longitudinally inhomogeneous prominence thread model.

 This paper is organized as follows. Section~\ref{sec:model} contains the description of the prominence thread model and the numerical method used to solve the eigenvalue problem of kink oscillations. The results of the computations are given in Section~\ref{sec:res}.  The implications of the findings of this paper for prominence seismology are discussed in Section~\ref{sec:dis}. Finally, concluding remarks are given in Section~\ref{sec:con}.

\section{Prominence thread model and numerical solution of the eigenvalue problem}
\label{sec:model}

\begin{figure*}[!htp]
\centering
\includegraphics[width=1.5\columnwidth]{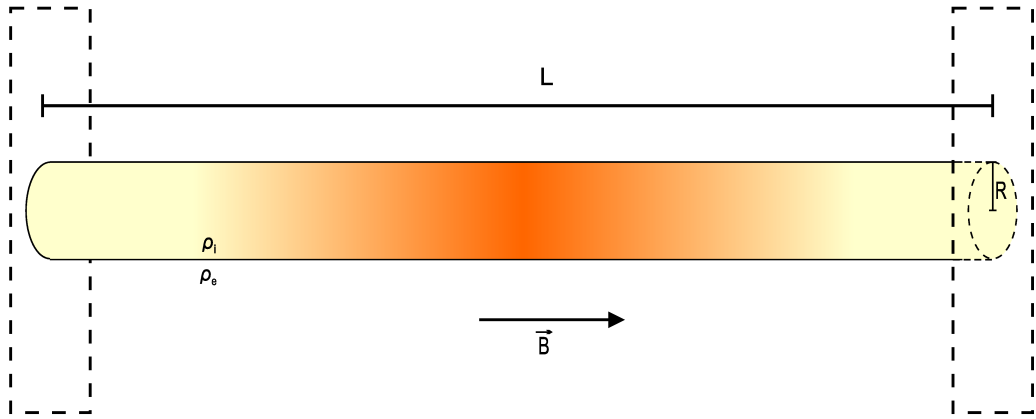}
\caption{Sketch of the prominence thread model used in this work. \label{fig:model}}
\end{figure*}

The equilibrium model of a prominence thread is schematically shown in Figure~\ref{fig:model}. We use a cylindrical coordinate system, with $r$, $\varphi$, and $z$ the radial, azimuthal, and longitudinal coordinates, respectively.  We consider a cylindrically symmetric straight magnetic flux tube of radius $R$ and length $L$.  The ends of the tube are located at $z=\pm L/2$ and are line-tied at two rigid walls representing the solar photosphere. The centre of the tube corresponds to $z=0$. The magnetic field, $\vec{B}$, is straight and along the axis of the tube. The magnetic field strength, $B_0$, is constant everywhere. We use the $\beta=0$ approximation, where $\beta$ refers to the ratio of the gas pressure to the magnetic pressure. This is an appropriate approximation to study kink oscillations, which are mainly driven by magnetic tension in thin tubes \citep{goossens2009}. In the $\beta=0$ approximation we can freely choose the spatial distribution of mass density. Hence, the equilibrium density, $\rho_0$, is
\begin{equation}
\rho_0(r,z) = \left\{ 
\begin{array}{lll}
\rhoi(z), &\textrm{if} & r \leq R, \\
\rhoe(z), &\textrm{if} & r > R,
\end{array} \right.
\end{equation}
where the internal, $\rho_{\rm i}(z)$, and external, $\rhoe(z)$, densities are functions of $z$ only, and $\rhoi(z) > \rhoe(z)$. The external plasma represents the coronal medium. In the present model the density jumps abruptly at the boundary of the tube, i.e., at $r=R$. We are aware that a continuous transition of density in the radial direction would be more realistic, although it would complicate matters. The kink oscillations of a transversely nonuniform tube  are resonantly coupled to Alfv\'en waves and, consequently, the  oscillations are damped in time \citep[see, e.g.,][]{goossens2011}. In such a case, the kink mode is no longer a true normal mode of the flux tube but a spectral pole of the Green's function, related to the initial-value problem, that represents a damped coordinated motion of the plasma \citep{sedlacek1971}. Since in the present work we are not interested in the damping, we assume a jump of density at the boundary of the tube to avoid the resonant absorption process. Readers are referred to \citet{arregui2008,arregui2011} and \citet{soler2009,soler2010} for more information about the resonant damping of prominence thread kink oscillations and to \citet{goossens2011} for a general review. A comparison of the efficiency of several damping mechanisms of transverse thread oscillations is given in \citet{soler2014}.

\begin{figure}[!htp]
\centering
\includegraphics[width=.85\columnwidth]{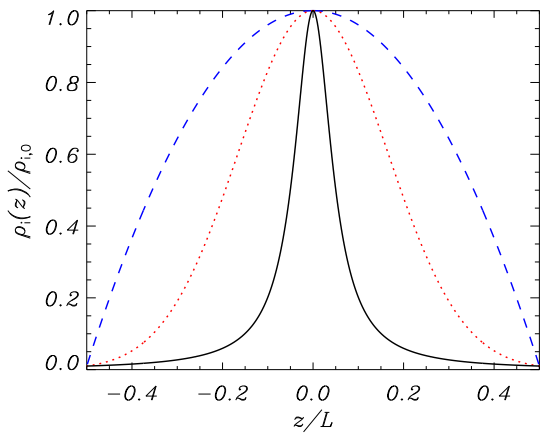}
\caption{Variation of the internal density along the prominence thread for the Lorentzian profile (black solid line), the Gaussian profile (red dotted line), and the parabolic profile (blue dashed line). We used $\chi = 100$ for representation purposes. \label{fig:profiles}}
\end{figure}

We denote the internal density at the centre and at the end of the flux tube as $\rhoi(0) = \rho_{\rm i,0}$ and $\rhoi(L/2) = \rho_{{\rm i},L/2}$, respectively. Then, we define the ratio of the two densities as $\chi = \rho_{\rm i,0} / \rho_{{\rm i},L/2}$, with $\chi \geq 1$. Contrary to coronal loops, prominence threads are denser near the centre of the thread than at their footpoints. The larger $\chi$, the stronger the density variation along the thread, with the case $\chi = 1$ corresponding to a homogeneous thread.  Here, we consider three different longitudinal profiles for the internal density, namely a Lorentzian profile
\begin{equation}
\rhoi(z) = \frac{\rho_{\rm i,0}}{ 1 + 4 \left( \chi - 1 \right) z^2/L^2},
\end{equation}
a Gaussian profile
\begin{equation}
\rhoi(z) = \rho_{\rm i,0} \exp \left[ - 4 \left( \log \chi \right) \frac{z^2}{L^2} \right],
\end{equation}
and a parabolic profile
\begin{equation}
\rhoi(z) = \rho_{\rm i,0} \left( 1 - 4 \frac{\chi - 1}{\chi} \frac{z^2}{L^2} \right).
\end{equation}
The three density profiles are compared in Figure~\ref{fig:profiles}. The three profiles represent different plasma arrangements within the magnetic tube. In the Lorentzian profile the densest plasma is narrowly concentrated near the centre of the tube, while the rest of the tube is occupied by much less dense plasma. In essence, the Lorentzian profile is similar to the piecewise  models used in previous works \citep[e.g.,][]{diaz2002,diaz2010,terradas2008,soler2010}. On the contrary, in the parabolic profile the dense prominence plasma is  broadly distributed along the tube.  The Gaussian profile represent an intermediate situation between the Lorentzian and parabolic profiles. The actual density variation along prominence threads is probably linked to the process that leads to the formation of  prominences and the condensation of the cool and dense plasma. A thermal instability might be involved in that process \citep[see, e.g.,][]{luna2012}. The purpose for choosing these three paradigmatic profiles is to determine the effect of different plasma arrangements on the period ratio of kink oscillations.

On the other hand, since we focus on horizontal threads, the external coronal plasma is assumed uniform for simplicity, namely $\rhoe(z) = \rho_{\rm e}$. Then, we define the prominence to corona density contrast  as the ratio of the internal density at the centre of the thread to the external density, namely $\zeta = \rho_{\rm i,0}/\rho_{\rm e}$.

The linear ideal MHD equations for a $\beta=0$ plasma that govern the behavior of small perturbations superimposed on the equilibrium prominence thread are 
\begin{eqnarray}
\rho_0 \frac{\partial \vec{v}}{\partial t} &=& \frac{1}{\mu} \left( \nabla \times \vec{b} \right)\times \vec{B}, \label{eq:basic1} \\
\frac{\partial \vec{b}}{\partial t} &=& \nabla \times \left( \vec{v} \times \vec{B} \right), \label{eq:basic2}
\end{eqnarray}
where $\vec{v}=(v_r,v_\varphi,v_z)$ and $\vec{b}=(b_r,b_\varphi,b_z)$ are the velocity and magnetic field perturbations, respectively, and $\mu$ is the magnetic permeability. Due to the $\beta=0$ approximation, $v_z=0$. Since  the equilibrium is invariant in the azimuthal direction,  we can put the perturbations proportional to $\exp \left( i m\varphi \right)$, where $m$ is the azimuthal wavenumber. We consider kink oscillations, so we set $m=1$. In addition, we perform a normal mode analysis and express the temporal dependence of the perturbations as $\exp \left( -i\omega t  \right)$, where $\omega$ is the  frequency. The period of the oscillation is $P=2\pi/\omega$. Then, Equations~(\ref{eq:basic1}) and (\ref{eq:basic2}) become 
\begin{eqnarray}
\omega v_r &=& \frac{B_0}{\mu \rho_0} \left( \frac{\partial b_r^*}{\partial z} -  \frac{\partial b_z^*}{\partial r} \right),  \label{eq:eigenini}\\
\omega v^*_\varphi &=& \frac{B_0}{\mu \rho_0} \left( \frac{b_z^*}{r} -  \frac{\partial b_\varphi}{\partial z} \right), \\
\omega b_r^* &=& - B_0 \frac{\partial v_r}{\partial z}, \\
\omega b_\varphi &=& B_0 \frac{\partial v_\varphi^*}{\partial z}, \\
\omega b_z^* &=& B_0 \left( \frac{\partial v_r}{\partial r} +  \frac{v_r}{r} + \frac{v_\varphi^*}{r} \right), \label{eq:eigenfin}
\end{eqnarray}
where,  with no loss of generality, we defined the new perturbations $v^*_\varphi = i v_\varphi$, $b_r^* = i b_r$, and $b_z^* = i b_z$ in order to remove complex numbers and so to have equations with real coefficients.  This is computationally convenient. Physically, the factor $i$ accounts for a phase difference of $\pi/2$. Therefore, the new perturbations $v_\varphi^*$, $b_r^*$, and $b_z^*$  are $\pi/2$ out of phase with respect to the actual perturbations $v_\varphi$, $b_r$, and $b_z$.  The remaining perturbations, namely $v_r$ and $b_\varphi$, are not altered. From here on, we drop the superscript * for simplicity. Equations~(\ref{eq:eigenini})--(\ref{eq:eigenfin}) define a  2D generalized eigenvalue problem, where $\omega$ is the eigenvalue and the perturbations form the eigenvector. Contrary to the piecewise constant models frequently used in the literature \citep[e.g.,][]{dymova2005,soler2010,diaz2010}, the analytic solution to the eigenvalue problem becomes very complicated when $\rho_0$ is an arbitrary function of $r$ and $z$ \citep[see][]{andries2005a}. In this work, the 2D eigenvalue problem  is numerically solved with the PDE2D code \citep{sewell}, a general-purpose partial differential equation solver. The PDE2D code was previously used to solve a similar 2D eigenvalue problem by \citet{arregui2011}.  The PDE2D code uses a collocation method, and the generalized matrix eigenvalue problem is solved using the shifted inverse power method. The code uses finite elements and allows the use of non-uniformly distributed grids. Different grid resolutions have been tested so as to assure the proper convergence of the solutions. The code provides the closest eigenvalue to an initially provided guess and the corresponding spatial form of the eigenfunctions.  The boundary conditions used in the code are those consistent with  trapped and standing kink oscillations. The two non-zero components of the velocity perturbation, namely $v_r$ and $v_\varphi$, and the longitudinal component of the  magnetic field perturbation, namely $b_z$, vanish at $z = \pm L/2$ because of the line-tying boundary condition at the photosphere. In turn, $b_r$ and $b_\varphi$ have vanishing longitudinal derivatives at $z = \pm L/2$. In the radial direction, $v_r$, $v_\varphi$, $b_r$ and $b_\varphi$ have vanishing radial derivatives and $b_z = 0$ at the axis of the tube, $r = 0$. The  code self-consistently connects the internal perturbations with the external perturbations at $r=R$ and satisfies the jump relations that naturally arise from the integration of the equations across $r=R$.  The condition that the modes are trapped imposes that all perturbations vanish far away from the tube in the radial direction, i.e.,  when $r \to \infty$. This condition is accomplished in the numerical code by setting all perturbations equal to zero at $r=r_{\rm max}$, where $r_{\rm max}$ is the largest value of the radial coordinate considered in the numerical domain. Ideally, we should set $r_{\rm max} \to \infty$. However, this is not  possible computationally, and $r_{\rm max}$ must be set to a finite value. We locate $r_{\rm max}$ sufficiently far from the tube boundary to properly compute the drop-off rate of perturbations in the radial direction and so avoid numerical errors.  To make sure that $r_{\rm max}$ is sufficiently large, we have performed convergence tests by increasing $r_{\rm max}$ until the solutions showed no dependence on this parameter. We took $r_{\rm max}=20 R$ in  the computations we include in this paper. In summary, the boundary conditions used in the numerical code are
\begin{eqnarray}
\frac{\partial v_r}{\partial r} = \frac{\partial v_\varphi}{\partial r} = \frac{\partial b_r}{\partial r} = \frac{\partial b_\varphi}{\partial r} =  0, \quad b_z = 0, \quad &\textrm{at}&  r=0, \\
 v_r =  v_\varphi =  b_r  =  b_\varphi = b_z = 0, \quad &\textrm{at}&  r=r_{\rm max}, \\
  v_r =  v_\varphi =  b_z = 0, \quad    \frac{\partial b_r}{\partial z} = \frac{\partial b_\varphi}{\partial z}= 0, \quad &\textrm{at}&  z=\pm L/2.
\end{eqnarray}

In the specific case that both the internal and external densities are uniform, namely $\rhoi(z) = \rho_{\rm i}$ and $\rhoe(z) = \rho_{\rm e}$,  it is possible to derive  analytic expressions for the period of kink oscillations in the thin tube (TT) limit, i.e., when $L/R \gg 1$ \citep[see, e.g.,][]{edwin1983,goossens2009}. The period of the fundamental longitudinal mode, $P_0$, and that of the first longitudinal overtone, $P_1$, are given by 
\begin{equation}
P_0 = \frac{2L}{\vai} \sqrt{\frac{1+\zeta}{2\zeta}}, \qquad P_1 = \frac{L}{\vai} \sqrt{\frac{1+\zeta}{2\zeta}}, \label{eq:ana}
\end{equation}
where $\vai = B_0/\sqrt{\mu \rho_{\rm i}}$ is the internal Alfv\'en velocity. Thus, the period ratio is $P_0/P_1 = 2$ in longitudinally homogeneous thin tubes. In the present model, deviations from $P_0/P_1 = 2$ can be caused by dispersion  when the thickness of the thread does not satisfy the TT condition and by longitudinal variation of density. The departure from the TT limit is not likely to have a strong impact for prominence threads, since observations clearly show that threads are very thin and long structures \citep[see][]{lin2011}. Although only the part of the tubes filled with the densest and coolest plasma can be seen in, e.g., H$\alpha$ and Ca II observations, the length of the whole magnetic tube, $L$, must be much longer than the observed length of the prominence threads, $L_{\rm obs}$. The values of $R$ and $L_{\rm obs}$ reported by the observations \citep[e.g.,][]{okamoto2007,lin2008,lin2011} are in the ranges 50~km~$\lesssim R \lesssim $~300~km and 3000~km~$\lesssim L_{\rm obs} \lesssim$~28,000~km, respectively, while \citet{okamoto2007} estimated $L$ to be at least $L\sim 10^5$~km. These numbers give  $10 \lesssim L_{\rm obs}/R \lesssim 400$ and $400 \lesssim L/R \lesssim 2000$, meaning that real prominence threads are actually very thin tubes.  Therefore, in the remainder of this paper we shall focus on the effect of longitudinal density variation, which may have a much stronger impact on the period ratio.

\section{Results}
\label{sec:res}

 \begin{figure}[!htp]
\centering
\includegraphics[width=.85\columnwidth]{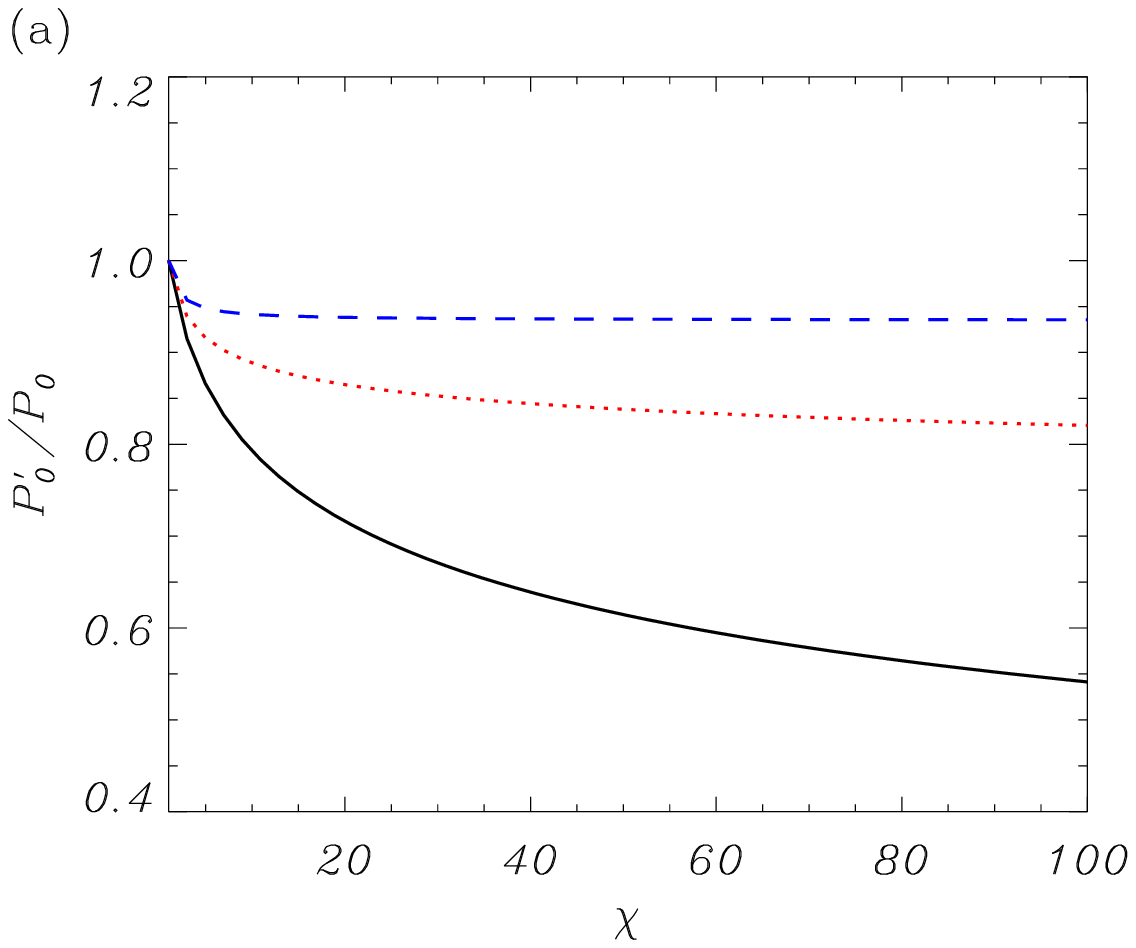}\\
\includegraphics[width=.85\columnwidth]{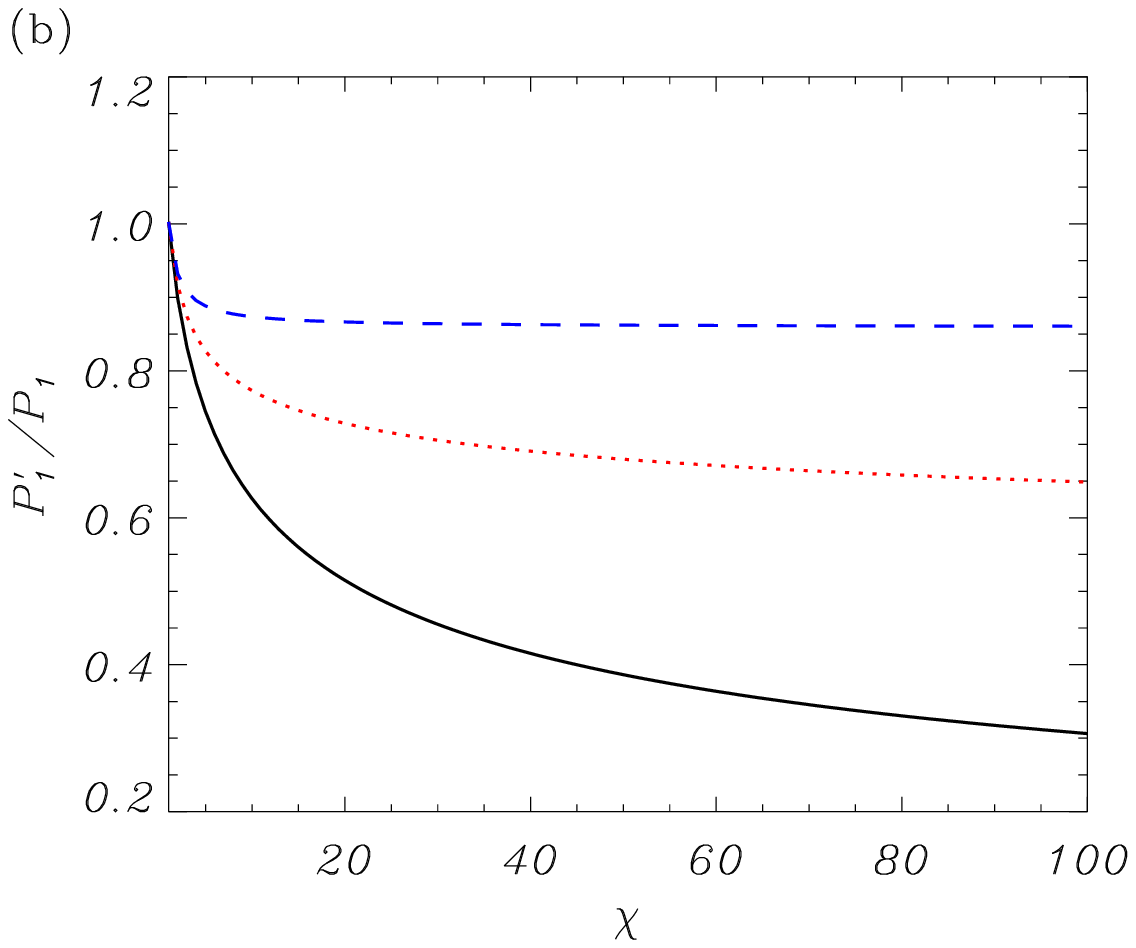}\\
\includegraphics[width=.85\columnwidth]{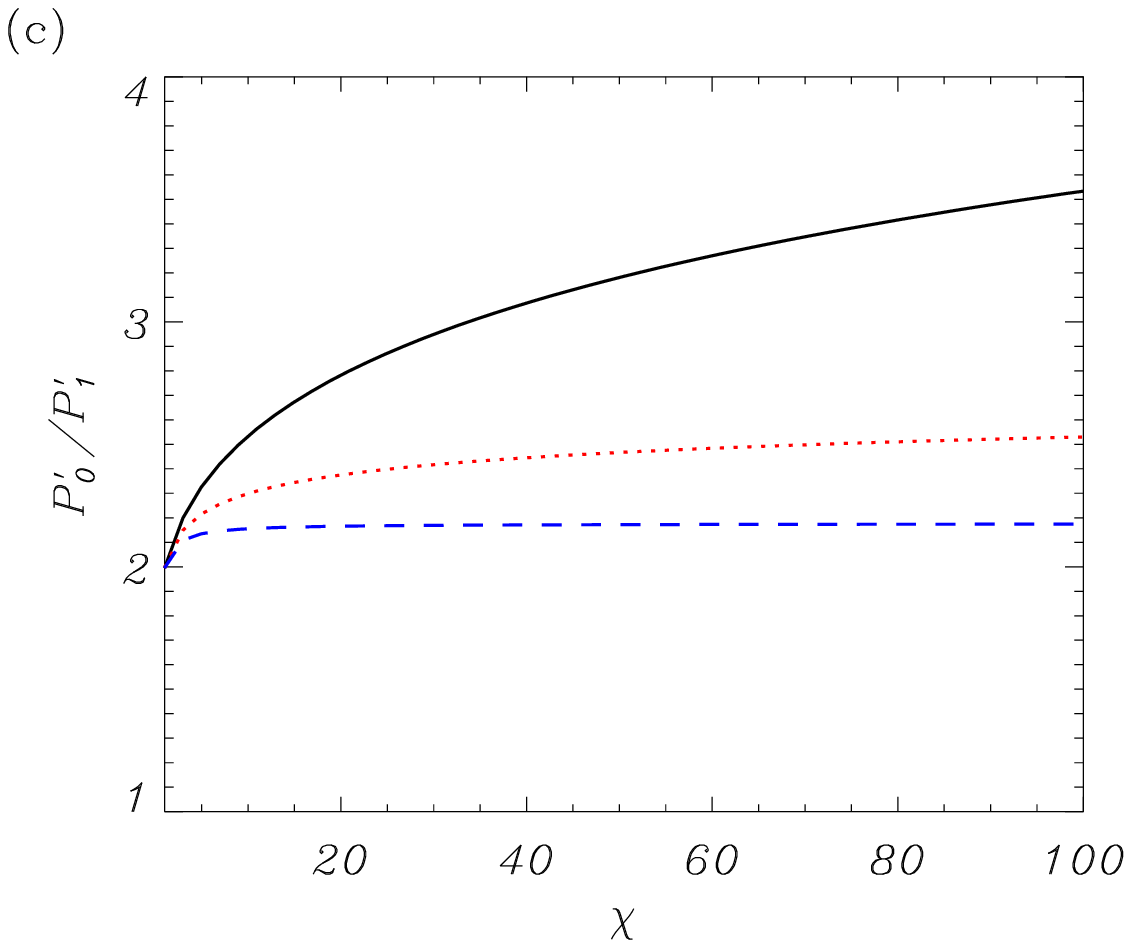}
\caption{Dependence of the numerically computed (a) fundamental mode period, $P_0'$,  (b) first overtone period, $P_1'$, and (c) period ratio, , $P_0'/P_1'$, with $\chi$ for the Lorentzian profile (black solid line), the Gaussian profile (red dotted line), and the parabolic profile (blue dashed line). We used $L/R = 100$ and $\zeta=100$ in all cases. $P_0$ and $P_1$ denote the periods of a homogeneous thread with $\rhoi = \rho_{\rm i,0}$. \label{fig:res1}}
\end{figure}

 \begin{figure}[!htp]
\centering
\includegraphics[width=.85\columnwidth]{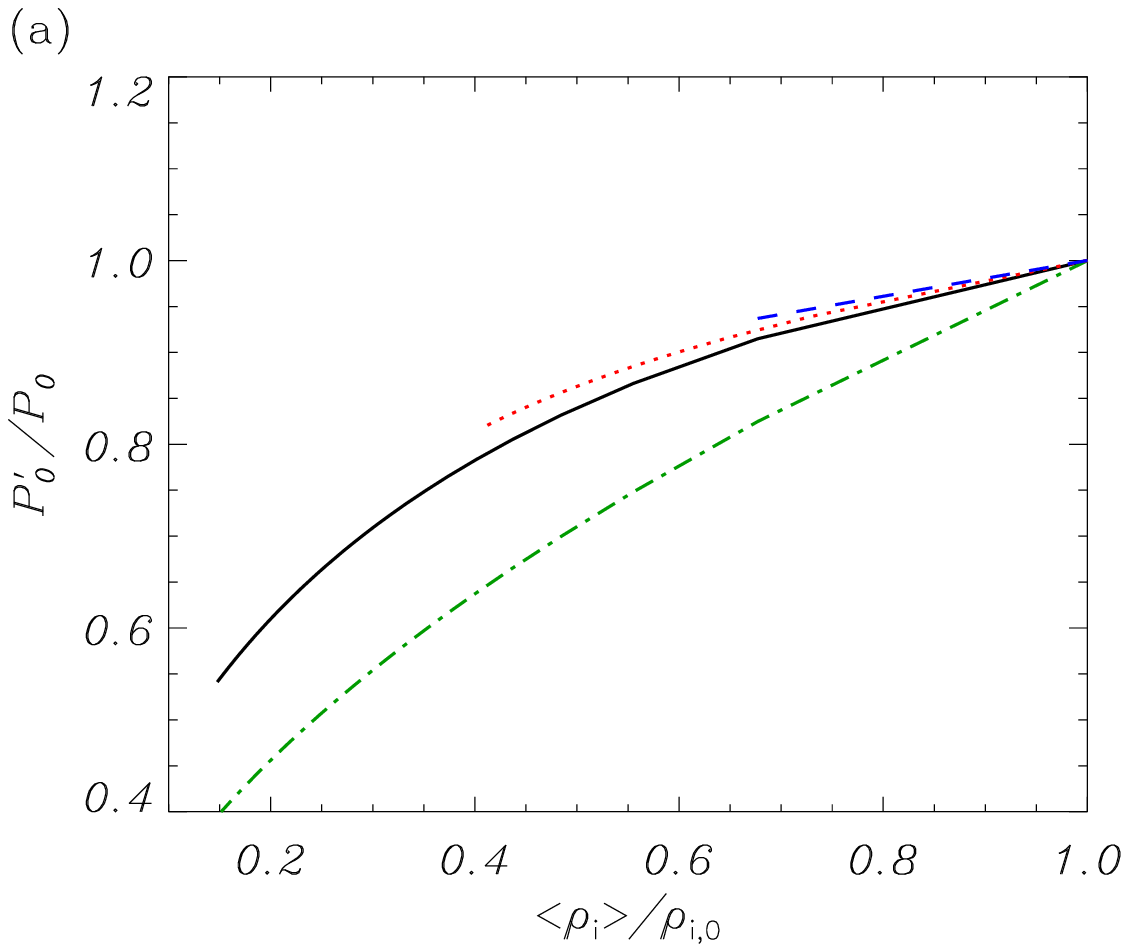}\\
\includegraphics[width=.85\columnwidth]{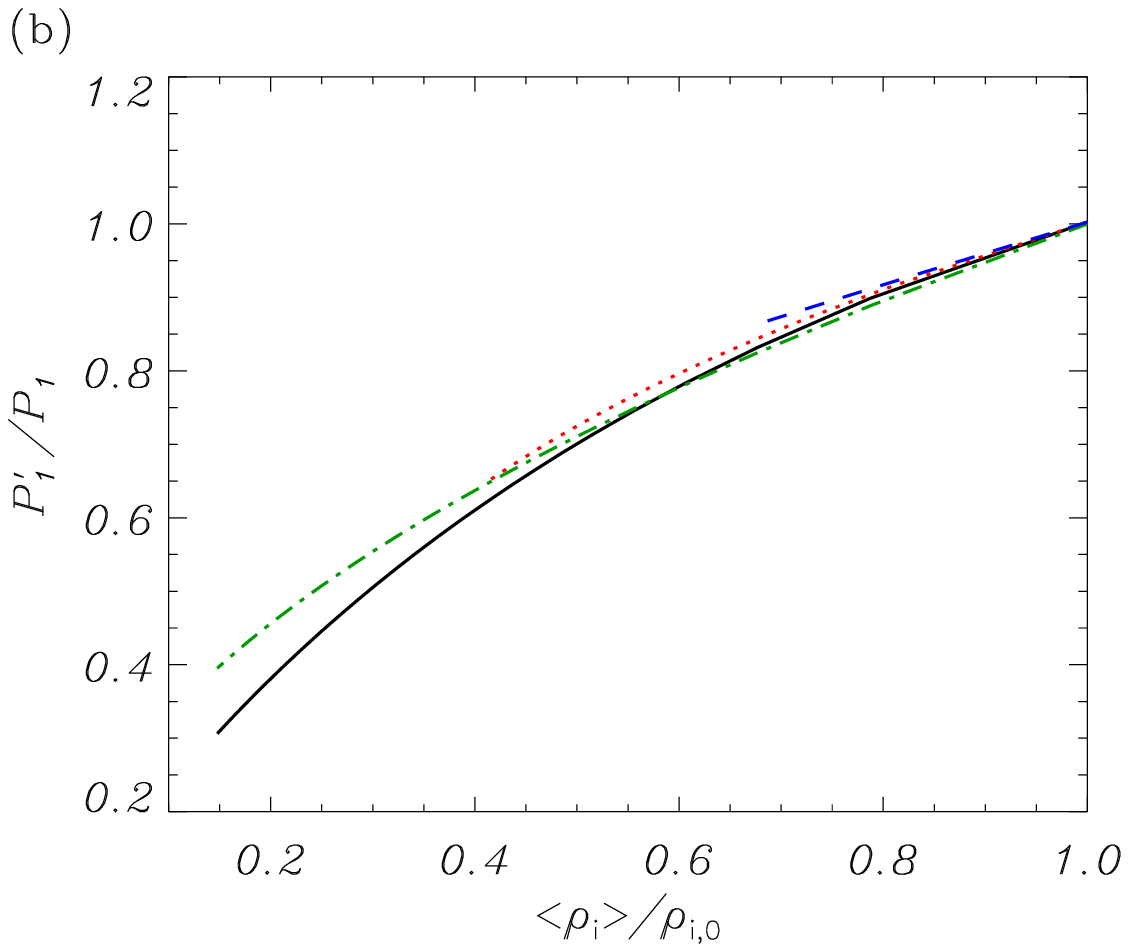}\\
\includegraphics[width=.85\columnwidth]{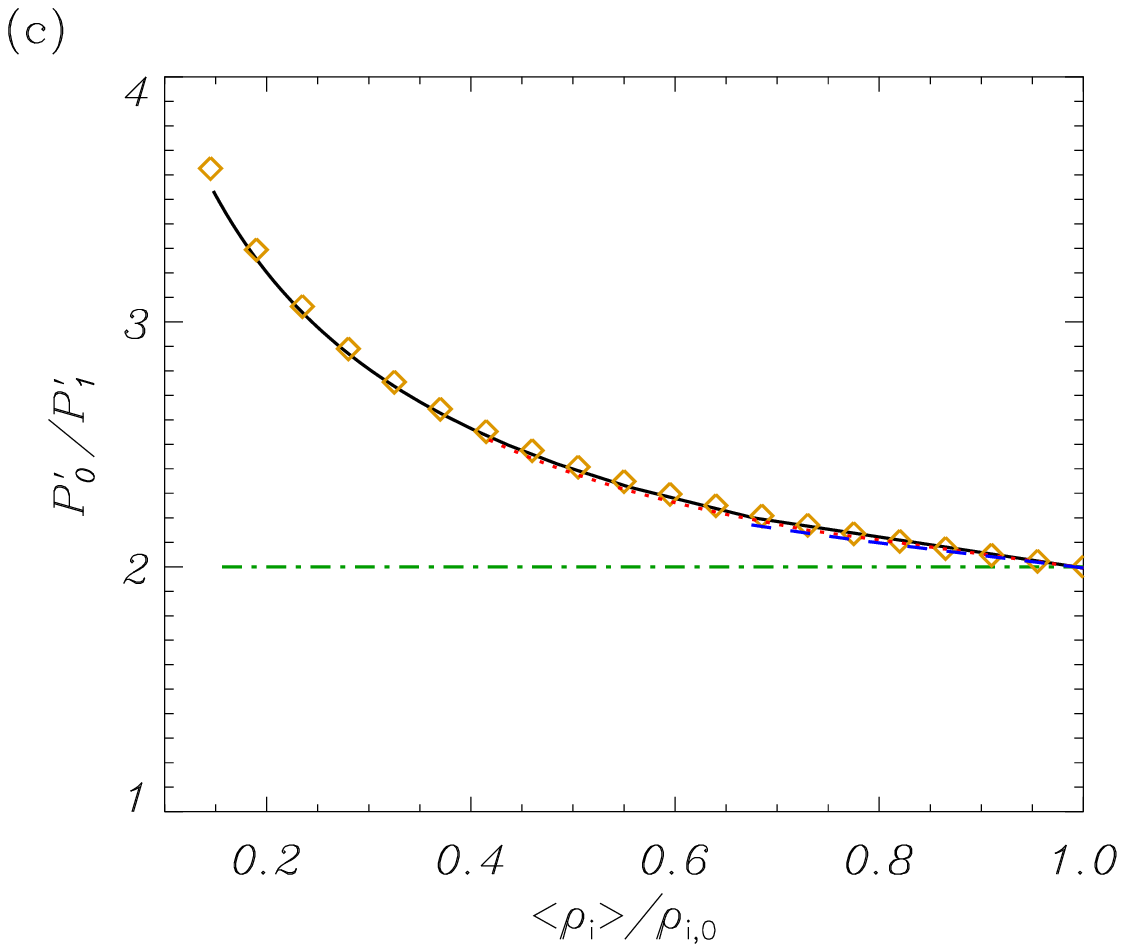}
\caption{Same results as Figure~\ref{fig:res1} but as function of the ratio of the average internal density to the central density, $\left< \rho_{\rm i} \right>/\rho_{\rm i,0}$. For comparison, the green dashed-dotted line shows the result for a homogeneous thread with $\rhoi = \left< \rhoi \right>$, i.e., the `average thread'. The symbols in panel (c) correspond to the empirical fit of Equation~(\ref{eq:fit}).  \label{fig:res2}}
\end{figure}

We start by investigating the modification of the fundamental mode and first overtone periods when the ratio of the central density to footpoint density, $\chi$, is increased from $\chi = 1$ (homogeneous tube) to $\chi = 100$. This is achieved by reducing the density at the footpoints while the central density is kept fixed. These results are displayed in Figure~\ref{fig:res1}. We denote by $P_0'$ and $P_1'$ the fundamental mode and first overtone periods, respectively, numerically computed with the PDE2D code. In turn, $P_0$ and $P_1$ are the periods of a homogeneous tube with $\rhoi = \rho_{\rm i,0}$ computed from Equation~(\ref{eq:ana}).

When $\chi$ increases, both the fundamental mode (Figure~\ref{fig:res1}a) and first overtone (Figure~\ref{fig:res1}b) periods decrease with respect to the values for a homogeneous tube with a density equal to the central density since, in average, the internal density is reduced. The decrease of the periods depends on the specific density profile considered. In this regard, the parabolic profile produces the smallest decrease of the periods. The results for the parabolic profile saturate at $\chi \approx 10$ and become independent of $\chi$ afterwards. On the contrary, the Lorentzian profile produces the largest decrease of the periods, which keep decreasing as $\chi$ increases. The results for the Gaussian profile are in between those of the  Lorentzian and parabolic profiles.

Figure~\ref{fig:res1}c shows the dependence with $\chi$ of the numerically computed period ratio $P_0'/P_1'$. Consistently, $P_0'/P_1' = 2$ when $\chi=1$. As $\chi$ increases, $P_0'/P_1'$ takes values larger than 2. This result conceptually agrees with that obtained by \citet{diaz2010} in tubes with  a piecewise constant density profile. Here, we find that $P_0'/P_1' > 2$ in the case of continuous density profiles as well. Again, how large is the increase of the period ratio from 2 depends strongly on the density profile considered. The Lorentzian profile produces a significant increase of the period ratio, so that $P_0'/P_1' \approx 3.5$ when $\chi = 100$. Conversely, the increase of the period ratio is small for the parabolic profile and  $P_0'/P_1' \approx 2.2$ when $\chi = 100$. Finally, we see again that the results for the Gaussian profile are in between those of the  Lorentzian and parabolic profiles, so that the period ratio increases moderately with $\chi$ for Gaussian profile and is $P_0'/P_1' \approx 2.5$ when $\chi = 100$. 

It should be noted that comparing the influence of the various profiles using the results of Figure~\ref{fig:res1} is not straightforward. When the ratio of the central density to footpoint density is fixed, the total mass in the thread is different for every profile. This fact makes  a direct comparison between profiles unsuitable. To fairly compare the results for the various profiles, the total mass in the thread should be the same, which is equivalent to considering the same average density in the thread \citep[see][]{andries2005}. Hence, we define the average internal density as
\begin{equation}
\left< \rho_{\rm i} \right> = \frac{1}{L} \int_{-L/2}^{L/2} \rhoi \left( z \right) {\rm d}z. \label{eq:average}
\end{equation}
Figure~\ref{fig:res2} displays the same results as Figure~\ref{fig:res1}, but now $\left< \rho_{\rm i} \right>/\rho_{\rm i,0}$ replaces $\chi$ in the horizontal axes of the plots. We see that there are no significant differences between the curves for the various profiles when the average internal density is the same. Importantly, Figure~\ref{fig:res2}c shows that the behavior of the period ratio is the same for the three profiles. The period ratio increases as $\left< \rho_{\rm i} \right>/\rho_{\rm i,0}$ decreases. This result indicates that the ratio of the average internal density to the central density, namely $\left< \rho_{\rm i} \right>/\rho_{\rm i,0}$, and not  the ratio of the central density to the footpoint density, namely $\chi$, is the parameter that actually controls the behavior of the period ratio regardless of the form of the density spatial variation along the thread \citep[see][for an equivalent result in the case of coronal loops]{andries2005}.  We  performed an empirical fit to the results of Figure~\ref{fig:res2}c and obtained that the dependence of the period ratio on the average density is very closely approximated by
\begin{equation}
\frac{P_0'}{P_1'} \approx 1 + \left( \frac{\left< \rho_{\rm i} \right>}{\rho_{\rm i,0}} \right)^{-1/2}, \label{eq:fit}
\end{equation}
for all the profiles. The empirical  Equation~(\ref{eq:fit}) is overploted in Figure~\ref{fig:res2}c to show that the fit is excellent. Equation~(\ref{eq:fit}) is independent of the density profile.  It is an important result that has simple and straightforward applications for prominence seismology. The implications of this result for prominence seismology are discussed in Section~\ref{sec:dis}.

\begin{figure}[!tb]
\centering
\includegraphics[width=.85\columnwidth]{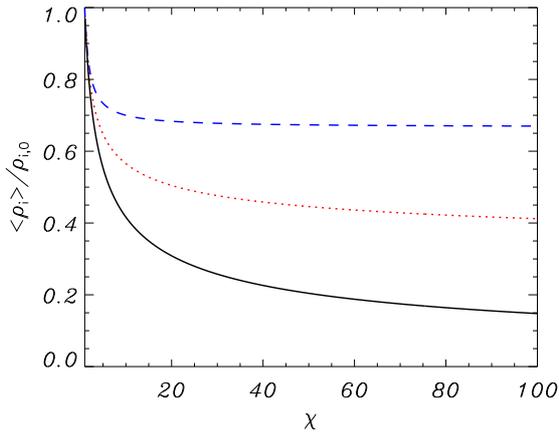}
\caption{Average internal density, $\left< \rho_{\rm i} \right>/\rho_{\rm i,0}$,  as a function of $\chi$. The various lines are for the Lorentzian profile (black solid line), the Gaussian profile (red dotted line), and the parabolic profile (blue dashed line). \label{fig:ave}}
\end{figure}

Note that the length of the curves in Figure~\ref{fig:res2} is different for different density profiles. The reason for this  is that $\left< \rho_{\rm i} \right>$ was computed by varying $\chi$ in the interval $[1,100]$, but the dependence of the average density on $\chi$ is different for every  profile.  The analytic result of computing the average density from Equation~(\ref{eq:average}) is
\begin{equation}
\left< \rho_{\rm i} \right> = \rho_{\rm i,0} \frac{\arctan\sqrt{\chi-1}}{\sqrt{\chi-1}},
\end{equation}
for the Lorentzian profile,
\begin{equation}
\left< \rho_{\rm i} \right> = \rho_{\rm i,0} \frac{\sqrt{\pi}}{2} \frac{{\rm erf}\sqrt{\log\chi}}{\sqrt{\log\chi}},
\end{equation}
for the Gaussian profile (where ${\rm erf}$ and $\log$ denote the error function and the natural logarithm, respectively), and
\begin{equation}
\left< \rho_{\rm i} \right> = \rho_{\rm i,0} \frac{2\chi + 1}{3\chi},
\end{equation}
for the parabolic profile. These average densities are plotted in Figure~\ref{fig:ave} as function of $\chi$. We see that the average density of the Lorentzian profile is lower than that of the parabolic profile for the same value of $\chi$, with that of the Gaussian profile in between. For prominence threads it is expected $\chi$ to be a large parameter. Hence, we perform the limit $\chi \gg 1$ and the expressions for the average density of the different profiles simplify to
\begin{equation}
\left< \rho_{\rm i} \right> \approx \rho_{\rm i,0} \frac{\pi}{2} \chi^{-1/2},
\end{equation}
for the Lorentzian profile,
\begin{equation}
\left< \rho_{\rm i} \right> \approx \rho_{\rm i,0} \frac{\sqrt{\pi}}{2} \left(\log\chi\right)^{-1/2},
\end{equation}
for the Gaussian profile, and
\begin{equation}
\left< \rho_{\rm i} \right> \approx \rho_{\rm i,0} \frac{2}{3},
\end{equation}
for the parabolic profile. Importantly, the average density of the parabolic profile becomes independent of $\chi$ when $\chi \gg 1$. Now we substitute these average densities into Equation~(\ref{eq:fit}) to obtain  simple relations between $P_0'/P_1'$ and $\chi$ for the various profiles, namely
\begin{equation}
\frac{P_0'}{P_1'} \approx 1 + \left( \frac{4}{\pi^2} \chi \right)^{1/4},\label{eq:lormax}
\end{equation}
for the Lorentzian profile,
\begin{equation}
\frac{P_0'}{P_1'} \approx 1 + \left( \frac{4}{\pi} \log \chi \right)^{1/4},\label{eq:gaumax}
\end{equation}
for the Gaussian profile, and
\begin{equation}
\frac{P_0'}{P_1'} \approx 1 + \sqrt{\frac{3}{2}} \approx 2.22, \label{eq:parmax}
\end{equation}
for the parabolic profile. These approximate results agree well with the numerical solutions displayed in Figure~\ref{fig:res1} when $\chi \gg 1$.

For comparison, we  overplot in Figure~\ref{fig:res2} the results for a longitudinally homogeneous thread whose internal density is equal to the average density, i.e., $\rhoi = \left< \rhoi \right>$. We call this hypothetical case the `average thread'.   Obviously, the period ratio of the `average thread' is 2 due to the absence of longitudinal inhomogeneity. However,  we see that the fundamental mode period of the `average thread' is shorter than the period of a nonuniform thread, but the first overtone periods are similar in both cases. The fact that the density varies spatially is important for the fundamental mode period, but is not very relevant for the first overtone period. This points out that ignoring the longitudinal variation of density in prominence threads  by using `average thread' models may provide inaccurate values of the fundamental mode period. This result is opposite to that found by \citet{andries2005} for coronal loops. \citet{andries2005} showed that the first overtone was more affected by plasma stratification than the fundamental mode. The reason for this difference is probably the fact that in prominence threads the densest plasma is located near the tube centre, while in stratified coronal loops the plasma is denser near the footpoints.  Also note that, contrary to  \citet{andries2005}, we do not perform a weighted spatial average of the density but an unweighed average.

 \begin{figure*}[!htp]
\centering
\includegraphics[width=.85\columnwidth]{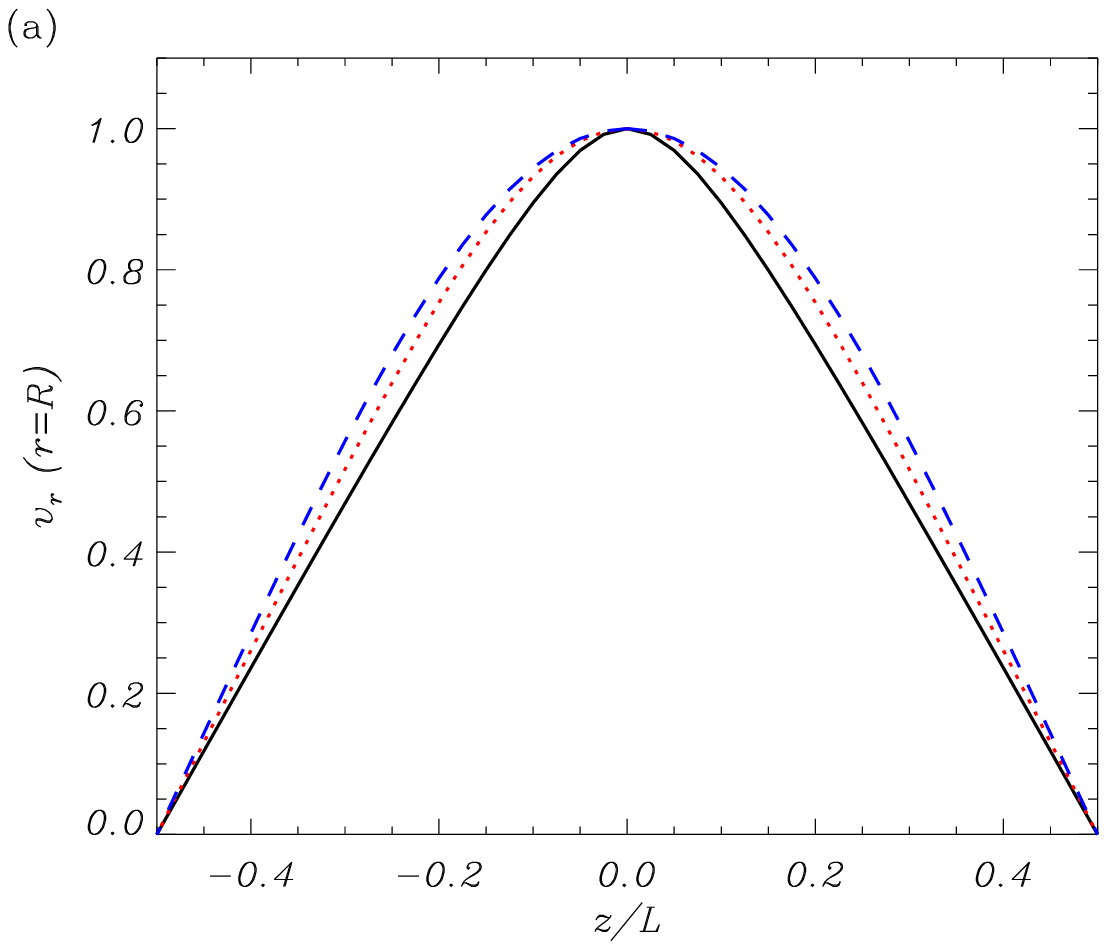} \includegraphics[width=.85\columnwidth]{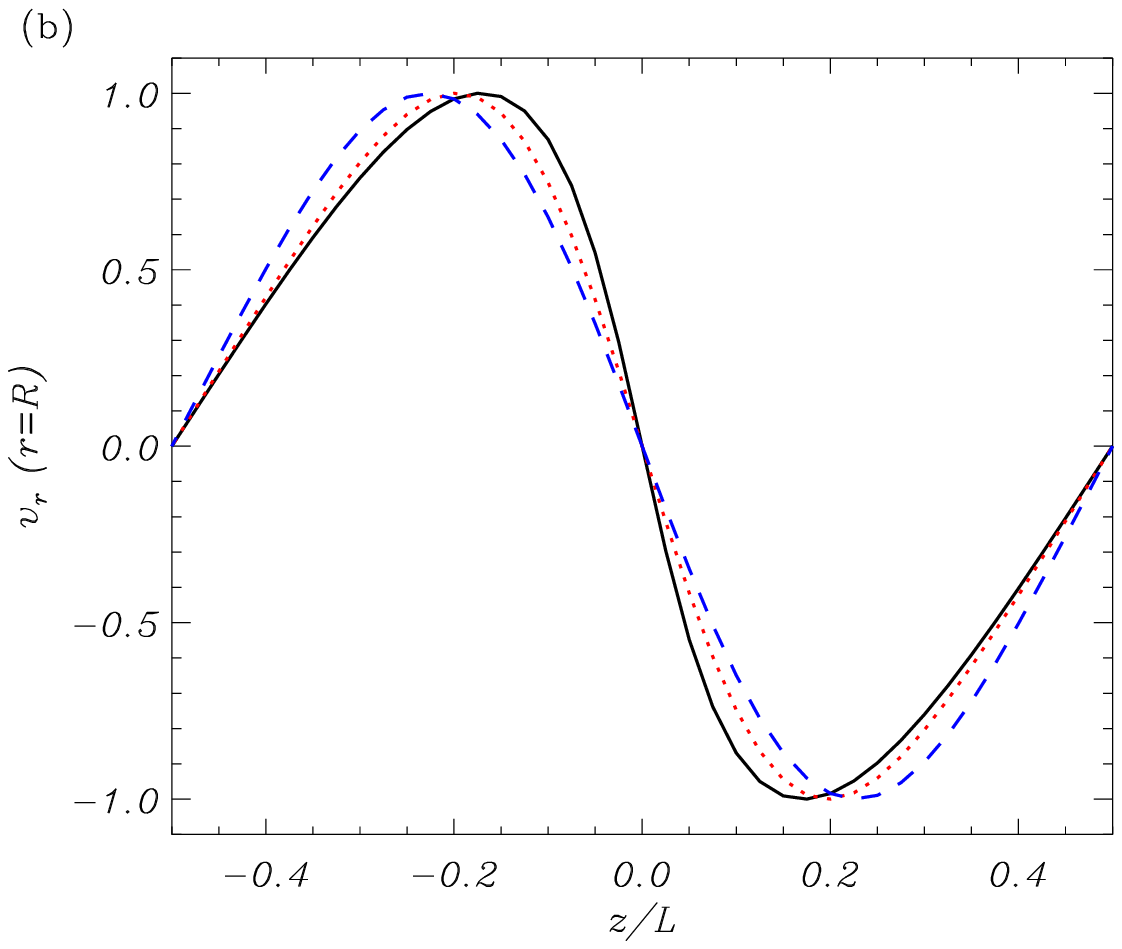}\\
\includegraphics[width=.85\columnwidth]{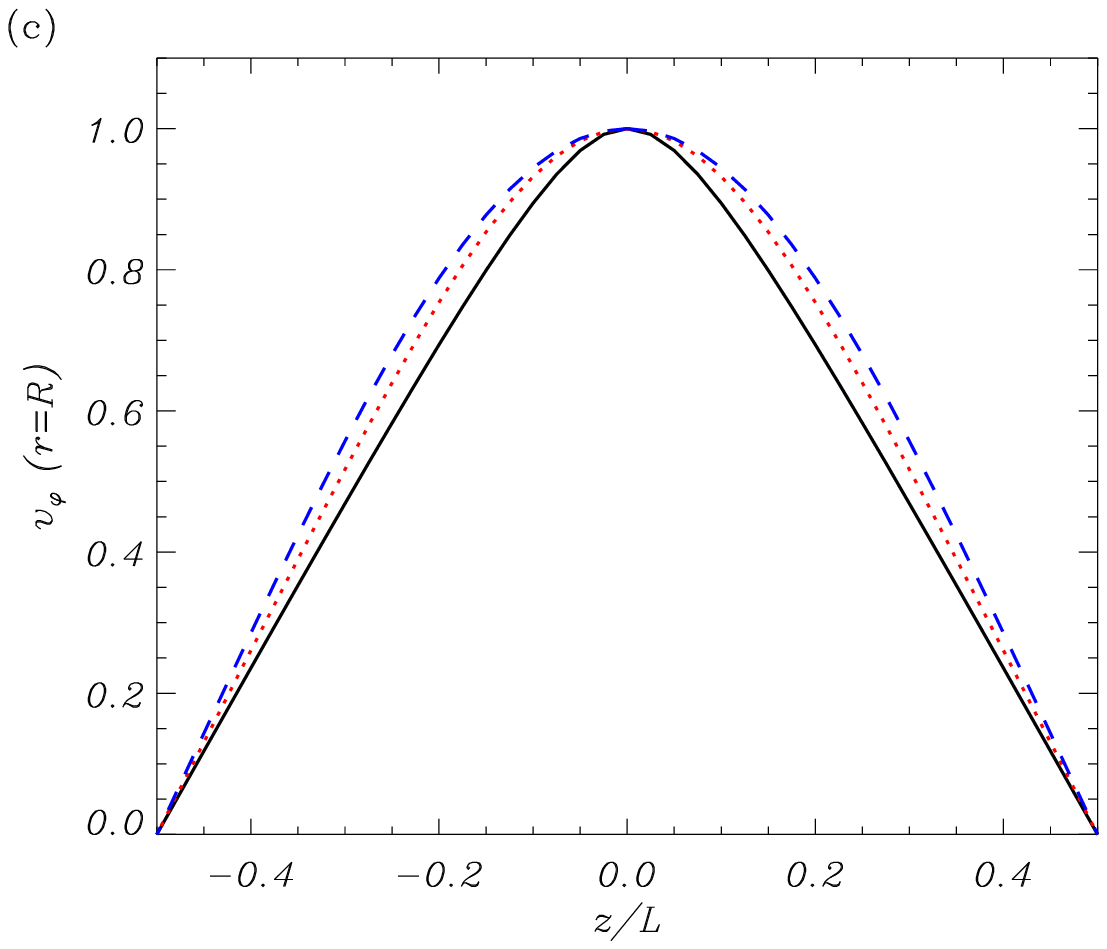} \includegraphics[width=.85\columnwidth]{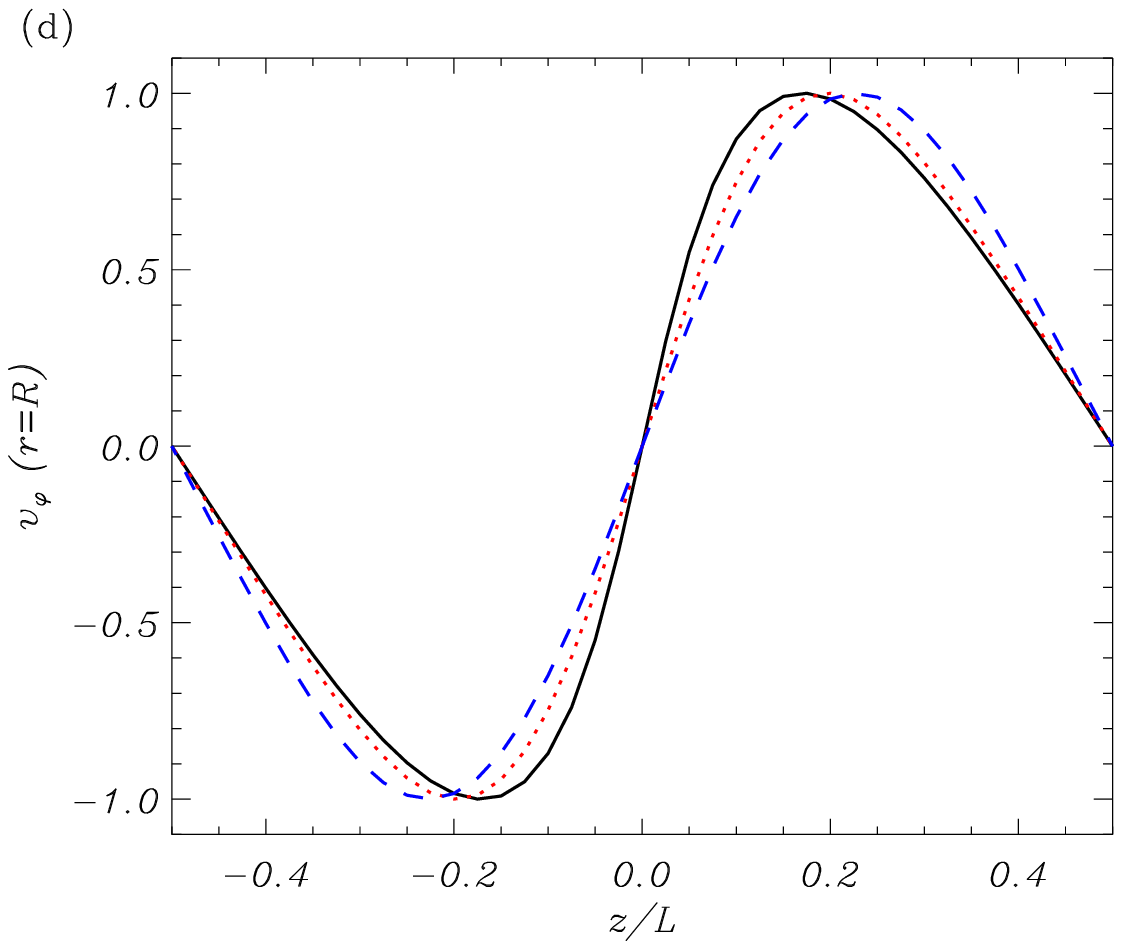}\\
\includegraphics[width=.85\columnwidth]{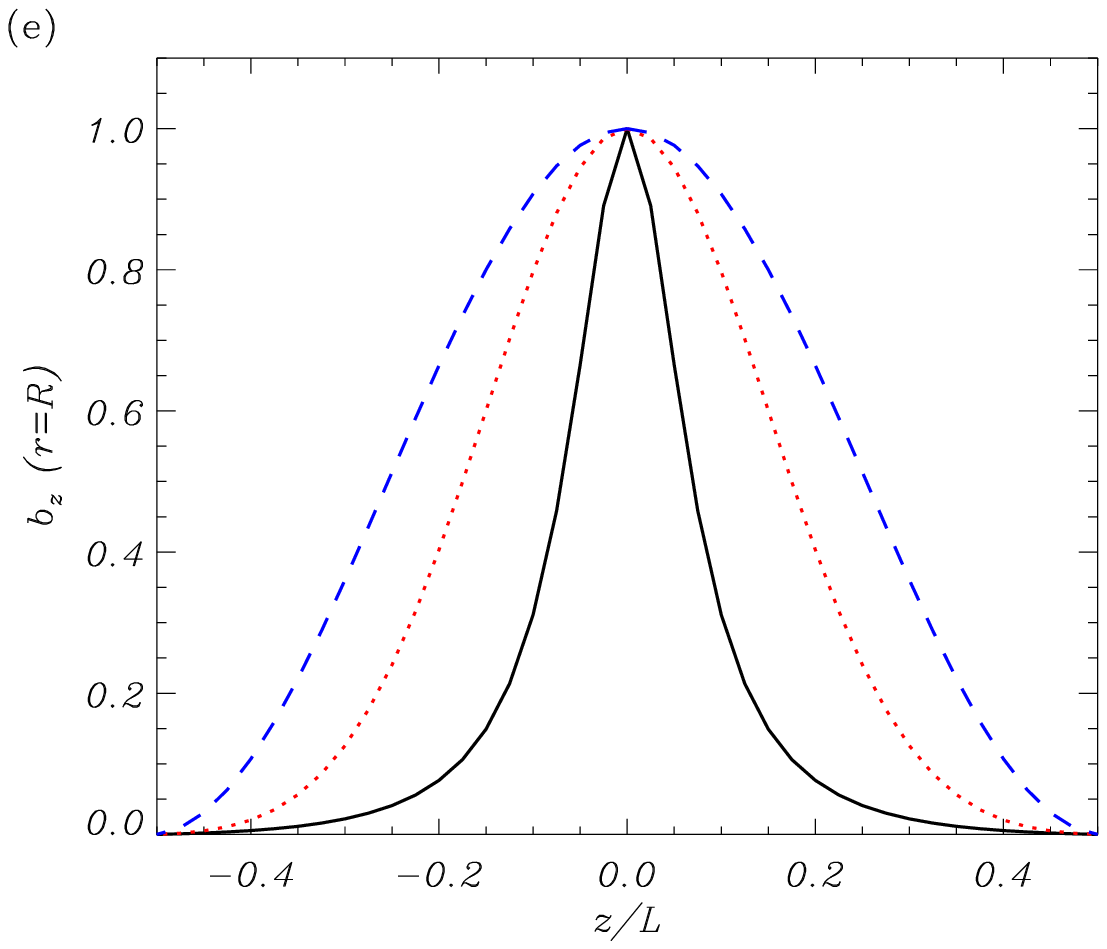} \includegraphics[width=.85\columnwidth]{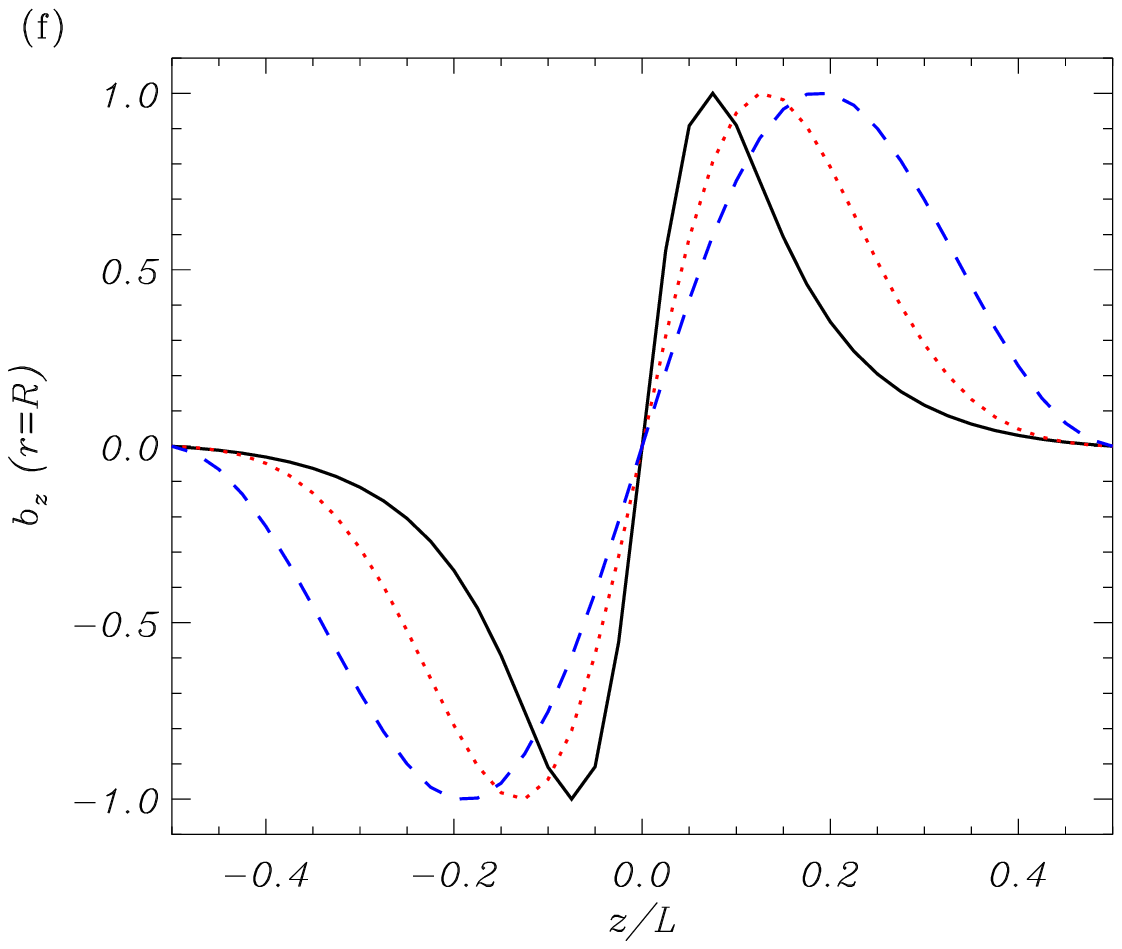}
\caption{Longitudinal cuts at $r=R$  of $v_r$ (top panels), $v_\varphi$ (mid panels), and $b_z$ (bottom panels) corresponding to the fundamental mode (left panels) and the first overtone (right panels). The various lines are for the Lorentzian profile (black solid line), the Gaussian profile (red dotted line), and the parabolic profile (blue dashed line). We used $L/R = 100$, $\chi=50$, and $\zeta=100$ in all cases. Arbitrary units are used. \label{fig:res3}}
\end{figure*}

Finally, for completeness we have computed the spatial form of the eigenfunctions with the PDE2D code. We display in Figure~\ref{fig:res3} longitudinal cuts of $v_r$, $v_\varphi$, and $b_z$ near the boundary of the thread, i.e., $r\approx R$, for both the fundamental mode and the first overtone using a particular set of parameters. Figure~\ref{fig:res3} may be compared with the eigenfunctions in piecewise-like models \citep{soler2010,arregui2011}. We can observe in Figure~\ref{fig:res3} that the eigenfunctions of the fundamental longitudinal mode of transverse kink waves (left panels) have no nodes in the axial (longitudinal) direction, while the eigenfunctions of the first longitudinal overtone of transverse kink waves (right panels) have one node at the centre of the tube. We find that  $v_r$ and $v_\varphi$  corresponding to the different profiles are very similar for both fundamental mode and first overtone. This  suggests  that measurements of the thread transverse velocity or, equivalent, the displacement could provide little information about the density profile. However, the spatial form of  $b_z$ is sensitive to the density profile. The more concentrated the dense prominence plasma near the centre of the tube, the faster the amplitude of $b_z$ drops from the centre. This result was also noticed by \citet[][see their Figure~6c]{arregui2011}. In the present computations, $b_z$ is essentially confined near the centre of the thread for the Lorentzian profile, but it is almost a sinusoidal-type function for the parabolic profile.

\section{Prominence seismology}
\label{sec:dis}

The results of this paper and, in particular, the empirical fit of Equation~(\ref{eq:fit}) have important implications for seismology of prominences. Equation~(\ref{eq:fit}) is general and does not depend on the specific density profile. This means that it can be used even when the density profile is not known. By inverting Equation~(\ref{eq:fit}), we find that the ratio of the average internal density to the central density is related to the period ratio as
\begin{equation}
\frac{\left< \rhoi \right>}{\rho_{\rm i,0}} \approx \left( \frac{P_0'}{P_1'}  - 1 \right)^{-2}. \label{eq:seis}
\end{equation}
Equation~(\ref{eq:seis}) can be used to seismologically estimate the ratio $\left< \rho_{\rm i} \right>/\rho_{\rm i,0}$ in prominence threads using observations of the period ratio. 

As mentioned in the Introduction, no reliable simultaneous observations of the two periods in prominence threads are currently available. Two periods were reported  by \citet{lin2007}. These authors studied an extensive area in a prominence and it is unclear whether the two periods  originate from the same thread or correspond to different threads.  However, to perform a simple exemplifying application, let us proceed as in \citet{diaz2010} and assume that the two periods reported by \citet{lin2007} actually correspond to the fundamental mode and the first overtone of the same oscillating thread. The two periods are $P_0' =$~16~min and $P_1' =$~3.6~min, which give $P_0'/P_1' \approx 4.44$. Then, using Equation~(\ref{eq:seis}) we get $\left< \rho_{\rm i} \right>/\rho_{\rm i,0} \approx 0.084$. This result indicates that there is a strong density gradient along the thread, since the average density is a small fraction of the central density. This is compatible with a large value of the ratio of the central density to the footpoint density, i.e.,  $\chi \gg 1$.

Next, we can also use the observed  $P_0'/P_1'$ to  infer the ratio of the central density to the footpoint density, $\chi$. To do so, we must necessarily assume a particular density profile along the thread. Note that the form of the density profile was not needed to estimate $\left< \rho_{\rm i} \right>/\rho_{\rm i,0}$. In the case of the Lorentzian profile we get from Equation~(\ref{eq:lormax}) that
\begin{equation}
\chi \approx \frac{\pi^2}{4} \left( \frac{P_0'}{P_1'}  - 1 \right)^4 \approx 347,
\end{equation}
while for the Gaussian profile we get from Equation~(\ref{eq:gaumax}) that
\begin{equation}
\chi \approx \exp\left[ \frac{\pi}{4} \left( \frac{P_0'}{P_1'}  - 1 \right)^4 \right] \approx 10^{48}.
\end{equation}
The value of $\chi$ corresponding to the Gaussian profile is unrealistically large, while the value corresponding to the Lorentzian profile is reasonably consistent with the expected densities in prominence threads. In the case of the parabolic profile, we found in  Equation~(\ref{eq:parmax}) that its period ratio is 2.22 when $\chi\gg 1$, so that the observed period ratio of 4.44 seems  hardly compatible with a parabolic profile. The inferred values of $\chi$ suggest that, among the three profiles studied here, the Lorentzian profile may provide the better explanation for the ratio of the two periods reported by \citet{lin2007}. However, a  rigorous comparison between models should be done and an adequate treatment of errors should be considered to obtain more robust conclusions \citep[see][]{arregui2013}.

\section{Concluding remarks}
\label{sec:con}

In this paper, we have investigated the ratio of the period of the fundamental mode to period of the first overtone of kink oscillations of longitudinally inhomogeneous prominence threads. Contrary to previous works that used piecewise constant density profiles \citep{diaz2010}, here we assumed a continuous nonuniform density profile along the thread, that is more consistent with the simulations of thread formation via plasma condensation in solar prominences \citep[e.g.,][]{luna2012}. Our results indicate that the ratio of the average internal density to the central density, namely $\left< \rho_{\rm i} \right>/\rho_{\rm i,0}$,  is the parameter that controls the behavior of the period ratio regardless of the form of the density spatial variation along the thread. A fit of the period ratio as a function of $\left< \rho_{\rm i} \right>/\rho_{\rm i,0}$ was obtained, which can be used to estimate the degree of nonuniformity along the threads if observations of the two periods are reported. From the observational point of view, the detection of the first overtone is challenging, since the amplitude in the densest part of the tube is small for that mode. Although difficult, we believe that the observation could be done with present or future high-resolution instruments. For instance, using recent observations with the Hinode satellite, \citet{hillier2013} reported a large number of oscillations in vertical prominence threads with velocity amplitudes as low as 0.2~km~s$^{-1}$. In view of these observations, the detection of oscillations with such small amplitudes in horizontal threads is more than likely.

Here we have studied the period ratio from a purely numerical point of view. Hence, the equation that relates the period ratio with the average density (Equation~(\ref{eq:fit})) was empirically obtained from the numerical data. Although it is beyond the purpose of the present study, it would be desirable that future works pursue a more robust theoretical, i.e., analytical, explanation for this relation. The analytical theory developed by \citet{dymova2005} could be used to do so.

In the model, we assumed for simplicity that the density within the flux tube is symmetric with respect to $z=0$. The effect of shifting the density enhancement from the tube centre was investigated in a previous paper by \cite{soler2010} using a simple piecewise model. The conclusion from \citet{soler2010} was that the periods of kink oscillations are weakly affected by the position of the densest plasma within the tube unless the density enhancement is put very close to the footpoints, which is not realistic in the case of horizontal prominence threads. Therefore, we expect that the use of an asymmetric density profile would not change significantly the results of the present paper.

The results of the seismology application of Section~\ref{sec:dis} suggest that value of the period ratio could be used to distinguish between different density profiles. In a general sense, rigorous statistical methods to perform model comparison, as the Bayesian analysis used by \citet{arregui2013}, could be very useful to determine what kind of density variations are the most probable ones for a given value of the period ratio. Future works should exploit this possibility. Also, the presence of uncertainties in the measurements of the periods should be properly accounted for.

Finally, we should mention that there are some effects not included in the prominence thread model that may influence the value of the period ratio. For instance, an ingredient missing from the model is the presence of mass flows along the thread. Simultaneous transverse oscillations and flows in prominence fine structures have been observed \citep[e.g.,][]{okamoto2007}. The effect of flows on the period ratio has been investigated in piecewise constant models \citep[e.g.,][]{soler2011,robertus2014}, but not in threads with nonuniform density profiles. Among other effects, the roles of magnetic field twist, curvature, and cross section  have been studied in the case of coronal loops \citep[see, e.g.,][]{verth2008,verthetal2008,ruderman2008,morton2009,karami2012} but they should be explored for prominence threads as well. In addition, in view of the large amount of observations \citep[see, e.g.,][]{hillier2013}, the problem of the transverse oscillations of vertical threads is also theoretically relevant. For vertical threads, the effect of gravitational stratification should be included in the model. A comparison between the properties of the oscillations of vertical and horizontal threads is an interesting task to carry out in forthcoming works.

\begin{acknowledgements}
We thank I\~nigo Arregui for reading a draft of this paper and for giving helpful comments. We also thank Teimuraz Zaqarashvili for  discussions on the \citet{lom2014} paper. We acknowledge support from MINECO and FEDER funds through project AYA2011-22846. RS also acknowledges support from CAIB through the `Grups Competitius' program and FEDER funds, from MINECO through a `Juan de la Cierva' grant, from MECD through project CEF11-0012, and from the `Vicerectorat d'Investigaci\'o i Postgrau' of the UIB. M.G. acknowledges support from KU Leuven via GOA/2009-009 and also partial support from the Interuniversity Attraction Pols programme initiated by the Belgian Science Policy Office (IAP P7/08 Charm).
\end{acknowledgements}

\appendix

\section{Comment on the paper by Lomineishvili et al. (2014)}
\label{sec:lom}

The results computed in the present paper for a thread with a parabolic density profile are in disagreement with those obtained by \citet{lom2014} using the same model. \citet{lom2014} claim that the period ratio is 3, but we found that the correct value of the period ratio is 2.22, approximately. Here we explain the source of the discrepancy between the results of \citet{lom2014} and those of the present paper.

A property of the analysis by  \citet{lom2014} is the discontinuous behaviour of their solutions as a function of the inhomogeneity parameter $\alpha$ at $\alpha = 0$. The parameter $\alpha$ used by \citet{lom2014} is related to our parameter $\chi$ as $\alpha = \left( \chi - 1 \right)/\chi$, so that $\alpha=0$ corresponds to $\chi=1$ in our notation. As \citet{lom2014} observe (see the paragraph that follows their Equation~(19)) $\alpha = 0$ in their Equation~(17) leads to the solution of the homogeneous tube. They continue to say that after corresponding calculation, one may recover the well-known dispersion relation obtained by \citet{edwin1983}. The ratio of the period of the fundamental longitudinal mode to that of the first overtone for a homogeneous tube in the TT approximation is 2. For $\alpha \neq 0$, however small the value of $\alpha$ is, \citet{lom2014} claim that this ratio is 3 in the TT approximation. According to the analysis by \citet{lom2014} the period ratio varies in a discontinuous manner at $\alpha = 0$, where it jumps from 2 for $\alpha = 0$ to 3 for $\alpha > 0$. Since longitudinal density stratification is not a singular perturbation to the problem of linear waves on magnetic cylinders, this result is inconsistent.

Equations~(53) and (54) of \citet{lom2014} correspond to analytic expressions for the fundamental mode and first overtone frequencies in the TT approximation. Setting $\alpha\to 0$  in Equations~(53) and (54) of  \citet{lom2014} results in vanishing frequencies, i.e., infinite periods. This result can also be seen in the numerically obtained frequencies plotted in their Figure~3.  \citet{lom2014} do not recover the frequencies of  \citet{edwin1983} when $\alpha \to 0$.

According to \citet{zaqarashvili2014}, the modes obtained by  \citet{lom2014} owe their existence to inhomogeneity and so they disappear when the thread is homogeneous, i.e., when $\alpha \to 0$.  Also according to \citet{zaqarashvili2014},  \citet{lom2014} assumed open boundary conditions at the ends of the thread because in their model ``the thread itself probably is a part of much longer magnetic tube, which has a coronal density outside the thread''. However, fixed boundary conditions are necessary for standing kink oscillations.   Hence, the mathematical analysis of \citet{lom2014} does not lead to a dispersion relation valid for standing kink oscillations.

\bibliographystyle{aa.bst} 
\bibliography{refs}

\begin{thebibliography}{50}
\expandafter\ifx\csname natexlab\endcsname\relax\def\natexlab#1{#1}\fi

\bibitem[{{Andries} {et~al.}(2005{\natexlab{a}}){Andries}, {Arregui}, \&
  {Goossens}}]{andries2005}
{Andries}, J., {Arregui}, I., \& {Goossens}, M. 2005{\natexlab{a}}, \apjl, 624,
  L57

\bibitem[{{Andries} {et~al.}(2005{\natexlab{b}}){Andries}, {Goossens},
  {Hollweg}, {Arregui}, \& {Van Doorsselaere}}]{andries2005a}
{Andries}, J., {Goossens}, M., {Hollweg}, J.~V., {Arregui}, I., \& {Van
  Doorsselaere}, T. 2005{\natexlab{b}}, \aap, 430, 1109

\bibitem[{{Andries} {et~al.}(2009){Andries}, {van Doorsselaere}, {Roberts},
  {Verth}, {Verwichte}, \& {Erd{\'e}lyi}}]{andries2009}
{Andries}, J., {van Doorsselaere}, T., {Roberts}, B., {et~al.} 2009, \ssr, 149,
  3

\bibitem[{{Arregui} {et~al.}(2013){Arregui}, {Asensio Ramos}, \&
  {D{\'{\i}}az}}]{arregui2013}
{Arregui}, I., {Asensio Ramos}, A., \& {D{\'{\i}}az}, A.~J. 2013, \apjl, 765,
  L23

\bibitem[{{Arregui} {et~al.}(2012{\natexlab{a}}){Arregui}, {Ballester},
  {Oliver}, {Soler}, \& {Terradas}}]{arreguihinode2012}
{Arregui}, I., {Ballester}, J.~L., {Oliver}, R., {Soler}, R., \& {Terradas}, J.
  2012{\natexlab{a}}, in Astronomical Society of the Pacific Conference Series,
  Vol. 455, 4th Hinode Science Meeting: Unsolved Problems and Recent Insights,
  ed. L.~{Bellot Rubio}, F.~{Reale}, \& M.~{Carlsson}, 211

\bibitem[{{Arregui} {et~al.}(2012{\natexlab{b}}){Arregui}, {Oliver}, \&
  {Ballester}}]{arregui2012}
{Arregui}, I., {Oliver}, R., \& {Ballester}, J.~L. 2012{\natexlab{b}}, Living
  Reviews in Solar Physics, 9, 2

\bibitem[{{Arregui} {et~al.}(2011){Arregui}, {Soler}, {Ballester}, \&
  {Wright}}]{arregui2011}
{Arregui}, I., {Soler}, R., {Ballester}, J.~L., \& {Wright}, A.~N. 2011, \aap,
  533, A60

\bibitem[{{Arregui} {et~al.}(2008){Arregui}, {Terradas}, {Oliver}, \&
  {Ballester}}]{arregui2008}
{Arregui}, I., {Terradas}, J., {Oliver}, R., \& {Ballester}, J.~L. 2008, \apjl,
  682, L141

\bibitem[{{Ballester}(2014)}]{ballester2014}
{Ballester}, J.~L. 2014, in IAU Symposium, Vol. 300, IAU Symposium, 30--39

\bibitem[{{Berger} {et~al.}(2008){Berger}, {Shine}, {Slater}, {Tarbell},
  {Title}, {Okamoto}, {Ichimoto}, {Katsukawa}, {Suematsu}, {Tsuneta}, {Lites},
  \& {Shimizu}}]{berger2008}
{Berger}, T.~E., {Shine}, R.~A., {Slater}, G.~L., {et~al.} 2008, \apjl, 676,
  L89

\bibitem[{{Casini} {et~al.}(2003){Casini}, {L{\'o}pez Ariste}, {Tomczyk}, \&
  {Lites}}]{casini2003}
{Casini}, R., {L{\'o}pez Ariste}, A., {Tomczyk}, S., \& {Lites}, B.~W. 2003,
  \apjl, 598, L67

\bibitem[{{D{\'{\i}}az} {et~al.}(2002){D{\'{\i}}az}, {Oliver}, \&
  {Ballester}}]{diaz2002}
{D{\'{\i}}az}, A.~J., {Oliver}, R., \& {Ballester}, J.~L. 2002, \apj, 580, 550

\bibitem[{{D{\'{\i}}az} {et~al.}(2010){D{\'{\i}}az}, {Oliver}, \&
  {Ballester}}]{diaz2010}
{D{\'{\i}}az}, A.~J., {Oliver}, R., \& {Ballester}, J.~L. 2010, \apj, 725, 1742

\bibitem[{{Donnelly} {et~al.}(2007){Donnelly}, {D{\'{\i}}az}, \&
  {Roberts}}]{donnelly2007}
{Donnelly}, G.~R., {D{\'{\i}}az}, A.~J., \& {Roberts}, B. 2007, \aap, 471, 999

\bibitem[{{Dymova} \& {Ruderman}(2005)}]{dymova2005}
{Dymova}, M.~V. \& {Ruderman}, M.~S. 2005, \solphys, 229, 79

\bibitem[{{Dymova} \& {Ruderman}(2006)}]{dymova2006}
{Dymova}, M.~V. \& {Ruderman}, M.~S. 2006, \aap, 459, 241

\bibitem[{{Edwin} \& {Roberts}(1983)}]{edwin1983}
{Edwin}, P.~M. \& {Roberts}, B. 1983, \solphys, 88, 179

\bibitem[{{Erd{\'e}lyi} {et~al.}(2014){Erd{\'e}lyi}, {Hague}, \&
  {Nelson}}]{robertus2014}
{Erd{\'e}lyi}, R., {Hague}, A., \& {Nelson}, C.~J. 2014, \solphys, 289, 167

\bibitem[{{Goossens} {et~al.}(2011){Goossens}, {Erd{\'e}lyi}, \&
  {Ruderman}}]{goossens2011}
{Goossens}, M., {Erd{\'e}lyi}, R., \& {Ruderman}, M.~S. 2011, \ssr, 158, 289

\bibitem[{{Goossens} {et~al.}(2009){Goossens}, {Terradas}, {Andries},
  {Arregui}, \& {Ballester}}]{goossens2009}
{Goossens}, M., {Terradas}, J., {Andries}, J., {Arregui}, I., \& {Ballester},
  J.~L. 2009, \aap, 503, 213

\bibitem[{{Hillier} {et~al.}(2013){Hillier}, {Morton}, \&
  {Erd{\'e}lyi}}]{hillier2013}
{Hillier}, A., {Morton}, R.~J., \& {Erd{\'e}lyi}, R. 2013, \apjl, 779, L16

\bibitem[{{Karami} \& {Bahari}(2012)}]{karami2012}
{Karami}, K. \& {Bahari}, K. 2012, \apj, 757, 186

\bibitem[{{Lin}(2011)}]{lin2011}
{Lin}, Y. 2011, \ssr, 158, 237

\bibitem[{{Lin} {et~al.}(2007){Lin}, {Engvold}, {Rouppe van der Voort}, \& {van
  Noort}}]{lin2007}
{Lin}, Y., {Engvold}, O., {Rouppe van der Voort}, L.~H.~M., \& {van Noort}, M.
  2007, \solphys, 246, 65

\bibitem[{{Lin} {et~al.}(2008){Lin}, {Martin}, {Engvold}, {Rouppe van der
  Voort}, \& {van Noort}}]{lin2008}
{Lin}, Y., {Martin}, S.~F., {Engvold}, O., {Rouppe van der Voort}, L.~H.~M., \&
  {van Noort}, M. 2008, Advances in Space Research, 42, 803

\bibitem[{{Lin} {et~al.}(2009){Lin}, {Soler}, {Engvold}, {Ballester},
  {Langangen}, {Oliver}, \& {Rouppe van der Voort}}]{lin2009}
{Lin}, Y., {Soler}, R., {Engvold}, O., {et~al.} 2009, \apj, 704, 870

\bibitem[{{Lomineishvili} {et~al.}(2014){Lomineishvili}, {Zaqarashvili},
  {Zhelyazkov}, \& {Tevzadze}}]{lom2014}
{Lomineishvili}, S.~N., {Zaqarashvili}, T.~V., {Zhelyazkov}, I., \& {Tevzadze},
  A.~G. 2014, \aap, 565, A35

\bibitem[{{Luna} {et~al.}(2012){Luna}, {Karpen}, \& {DeVore}}]{luna2012}
{Luna}, M., {Karpen}, J.~T., \& {DeVore}, C.~R. 2012, \apj, 746, 30

\bibitem[{{McEwan} {et~al.}(2008){McEwan}, {D{\'{\i}}az}, \&
  {Roberts}}]{mcewan2008}
{McEwan}, M.~P., {D{\'{\i}}az}, A.~J., \& {Roberts}, B. 2008, \aap, 481, 819

\bibitem[{{McEwan} {et~al.}(2006){McEwan}, {Donnelly}, {D{\'{\i}}az}, \&
  {Roberts}}]{mcewan2006}
{McEwan}, M.~P., {Donnelly}, G.~R., {D{\'{\i}}az}, A.~J., \& {Roberts}, B.
  2006, \aap, 460, 893

\bibitem[{{Morton} \& {Erd{\'e}lyi}(2009)}]{morton2009}
{Morton}, R.~J. \& {Erd{\'e}lyi}, R. 2009, \aap, 502, 315

\bibitem[{{Ning} {et~al.}(2009){Ning}, {Cao}, {Okamoto}, {Ichimoto}, \&
  {Qu}}]{ning2009}
{Ning}, Z., {Cao}, W., {Okamoto}, T.~J., {Ichimoto}, K., \& {Qu}, Z.~Q. 2009,
  \aap, 499, 595

\bibitem[{{Okamoto} {et~al.}(2007){Okamoto}, {Tsuneta}, {Berger}, {Ichimoto},
  {Katsukawa}, {Lites}, {Nagata}, {Shibata}, {Shimizu}, {Shine}, {Suematsu},
  {Tarbell}, \& {Title}}]{okamoto2007}
{Okamoto}, T.~J., {Tsuneta}, S., {Berger}, T.~E., {et~al.} 2007, Science, 318,
  1577

\bibitem[{{Orozco Su{\'a}rez} {et~al.}(2014{\natexlab{a}}){Orozco Su{\'a}rez},
  {Asensio Ramos}, \& {Trujillo Bueno}}]{orozcofield2014}
{Orozco Su{\'a}rez}, D., {Asensio Ramos}, A., \& {Trujillo Bueno}, J.
  2014{\natexlab{a}}, \aap, in press

\bibitem[{{Orozco Su{\'a}rez} {et~al.}(2014{\natexlab{b}}){Orozco Su{\'a}rez},
  {D{\'{\i}}az}, {Asensio Ramos}, \& {Trujillo Bueno}}]{orozco2014}
{Orozco Su{\'a}rez}, D., {D{\'{\i}}az}, A.~J., {Asensio Ramos}, A., \&
  {Trujillo Bueno}, J. 2014{\natexlab{b}}, \apjl, 785, L10

\bibitem[{{Ruderman} {et~al.}(2008){Ruderman}, {Verth}, \&
  {Erd{\'e}lyi}}]{ruderman2008}
{Ruderman}, M.~S., {Verth}, G., \& {Erd{\'e}lyi}, R. 2008, \apj, 686, 694

\bibitem[{{Schmieder} {et~al.}(2010){Schmieder}, {Chandra}, {Berlicki}, \&
  {Mein}}]{schmieder2010}
{Schmieder}, B., {Chandra}, R., {Berlicki}, A., \& {Mein}, P. 2010, \aap, 514,
  A68

\bibitem[{{Sedl{\'a}{\v c}ek}(1971)}]{sedlacek1971}
{Sedl{\'a}{\v c}ek}, Z. 1971, Journal of Plasma Physics, 5, 239

\bibitem[{{Sewell}(2005)}]{sewell}
{Sewell}, G. 2005, {The Numerical Solution of Ordinary and Partial Differential
  Equations} (Wiley \& Sons)

\bibitem[{{Soler} {et~al.}(2010){Soler}, {Arregui}, {Oliver}, \&
  {Ballester}}]{soler2010}
{Soler}, R., {Arregui}, I., {Oliver}, R., \& {Ballester}, J.~L. 2010, \apj,
  722, 1778

\bibitem[{{Soler} \& {Goossens}(2011)}]{soler2011}
{Soler}, R. \& {Goossens}, M. 2011, \aap, 531, A167

\bibitem[{{Soler} {et~al.}(2014){Soler}, {Oliver}, \& {Ballester}}]{soler2014}
{Soler}, R., {Oliver}, R., \& {Ballester}, J.~L. 2014, in IAU Symposium, Vol.
  300, IAU Symposium, ed. B.~{Schmieder}, J.-M. {Malherbe}, \& S.~T. {Wu},
  48--51

\bibitem[{{Soler} {et~al.}(2009){Soler}, {Oliver}, {Ballester}, \&
  {Goossens}}]{soler2009}
{Soler}, R., {Oliver}, R., {Ballester}, J.~L., \& {Goossens}, M. 2009, \apjl,
  695, L166

\bibitem[{{Soler} {et~al.}(2012){Soler}, {Ruderman}, \& {Goossens}}]{soler2012}
{Soler}, R., {Ruderman}, M.~S., \& {Goossens}, M. 2012, \aap, 546, A82

\bibitem[{{Terradas} {et~al.}(2008){Terradas}, {Arregui}, {Oliver}, \&
  {Ballester}}]{terradas2008}
{Terradas}, J., {Arregui}, I., {Oliver}, R., \& {Ballester}, J.~L. 2008, \apjl,
  678, L153

\bibitem[{{Terradas} {et~al.}(2013){Terradas}, {Soler}, {D{\'{\i}}az},
  {Oliver}, \& {Ballester}}]{terradas2013}
{Terradas}, J., {Soler}, R., {D{\'{\i}}az}, A.~J., {Oliver}, R., \&
  {Ballester}, J.~L. 2013, \apj, 778, 49

\bibitem[{{Van Doorsselaere} {et~al.}(2007){Van Doorsselaere}, {Nakariakov}, \&
  {Verwichte}}]{vandoorsselaere2007}
{Van Doorsselaere}, T., {Nakariakov}, V.~M., \& {Verwichte}, E. 2007, \aap,
  473, 959

\bibitem[{{Verth} \& {Erd{\'e}lyi}(2008)}]{verth2008}
{Verth}, G. \& {Erd{\'e}lyi}, R. 2008, \aap, 486, 1015

\bibitem[{{Verth} {et~al.}(2008){Verth}, {Erd{\'e}lyi}, \&
  {Jess}}]{verthetal2008}
{Verth}, G., {Erd{\'e}lyi}, R., \& {Jess}, D.~B. 2008, \apjl, 687, L45

\bibitem[{{Zaqarashvili}(2014)}]{zaqarashvili2014}
{Zaqarashvili}, T.~V. 2014, private communication

\end{thebibliography}

\end{document}